\documentclass[preprint,aps,prd,showpacs,twocolumn,mamsmath,10pt]{revtex4}
\usepackage{graphicx}
\usepackage{epsfig,epsf,graphics,psfrag}
\usepackage{amsfonts,amssymb,stmaryrd,latexsym,amsmath}
\usepackage{enumerate}
\usepackage{slashed}

\newcommand{\be}{\begin{equation}}
\newcommand{\ee}{\end{equation}}
\newcommand{\bea}{\begin{eqnarray}}
\newcommand{\eea}{\end{eqnarray}}
\newcommand{\non}{\nonumber}

\newcommand{\lang}{\left\langle}
\newcommand{\rang}{\right\rangle}
\newcommand{\al}{&\!\!\!}

\begin{document}

\preprint{\scriptsize FZJ-IKP-TH-2010-03, HISKP-TH-10/05}
\title{ %\hfill{\tiny FZJ-IKP-TH-2010-03, HISKP-TH-10/05}\\[1.5em]
Novel analysis of the decays $\psi^\prime \to h_c
\pi^0$ and $\eta_c'\to \chi_{c0} \pi^0$}

\author{Feng-Kun Guo$^1$%\footnote{{\it E-mail address:} f.k.guo@fz-juelich.de}, %
       ~Christoph~Hanhart$^{1,2}$%\footnote{{\it E-mail address:} c.hanhart@fz-juelich.de},%
       ~Gang Li$^{3,4}$%\footnote{{\it E-mail address:} gli@ihep.ac.cn},\\%
       ~Ulf-G. Mei{\ss}ner$^{1,2,5}$%\footnote{{\it E-mail address:} meissner@itkp.uni-bonn.de},%
       ~and Qiang Zhao$^{3,6}$%\footnote{{\it E-mail address:} zhaoq@ihep.ac.cn}
       }

%\author{Feng-Kun Guo$^1$\footnote{{\it E-mail address:} f.k.guo@fz-juelich.de}
%, Christoph~Hanhart$^{1,2}$\footnote{{\it E-mail address:}
%c.hanhart@fz-juelich.de} , and Ulf-G. Mei{\ss}ner$^{1,2,3}$\footnote{{\it E-mail
%address:} meissner@itkp.uni-bonn.de} }

\affiliation{\small $\rm ^1$Institut f\"{u}r Kernphysik and J\"ulich Center
             for Hadron Physics, Forschungszentrum J\"{u}lich,
             D--52425 J\"{u}lich, Germany}%
\affiliation{\small $\rm ^2$Institute for Advanced Simulation,
             Forschungszentrum J\"{u}lich, D--52425 J\"{u}lich, Germany}%
\affiliation{\small $\rm ^3$Institute of High Energy Physics, Chinese Academy
             of Sciences, Beijing 100049, China}%
\affiliation{\small $\rm ^4$Department of Physics, Qufu Normal University, Qufu,
             273165, China}%
\affiliation{\small $\rm ^5$Helmholtz-Institut f\"ur Strahlen- und
             Kernphysik and Bethe Center for Theoretical Physics,\\ Universit\"at
             Bonn,  D--53115 Bonn, Germany}%
\affiliation{\small $\rm ^6$Theoretical Physics Center for Science Facilities,
             CAS, Beijing 100049, China}%

\begin{abstract}
\noindent We show that in the transitions $\psi^\prime \to h_c
\pi^0$ and $\eta_c'\to \chi_{c0} \pi^0$ the contributions from
charmed meson loops are highly suppressed, in contrast to various other
charmonium decays. We calculate the width of the $\psi^\prime \to h_c \pi^0$,
which agrees with the recent BES-III data, and predict the width of
the $\eta_c'\to \chi_{c0} \pi^0$, $\Gamma (\eta_c'\to \chi_{c0}
\pi^0) = 1.5\pm0.4$~keV. A confirmation of this prediction would
also provide additional support for a recent analysis
of $\psi'\to J/\psi \pi^0 (\eta)$, where loops are claimed
to play a prominent role.
\end{abstract}


\maketitle

\vspace{1cm}

\section{Introduction}

Recently, in a variety of calculations it was shown that charmed
meson loops contribute importantly to the decays of charmonia with
and without isospin breaking (for a overview, see \cite{Zhao:QNP}).
For instance, using an effective Lagrangian approach (ELA), the
intermediate meson loop contributions are found to be essential for
understanding the puzzling  $\psi(3770)$ non-$D\bar{D}$
decays~\cite{Zhang:2009kr,Liu:2009dr}. They are also important
in the $J/\psi$ decays into a vector and a pseudoscalar
mesons~\cite{Liu:2006dq} and in the $M1$ radiative transitions
between two charmonia~\cite{Li:2007xr}. For isospin breaking
decays, besides $\pi^0-\eta$ mixing and electromagnetic (e.m.)
effects, one also expects the mass difference between the neutral
and charged charmed mesons in the intermediate states (i.e. in the
meson loops) to play a role. This effect, known to
be of particular importance near the continuum thresholds,
was studied in the decays
 $\phi\to
\omega\pi^0$~\cite{Li:2008xm,Li:2007au},
$J/\psi\to \phi \eta \pi^0$~\cite{yourf0,ourf0},
and $D_{s0}^*(2317)\to
D_s\pi^0$~\cite{Faessler:2007gv,Lutz:2007sk,Guo:2008gp}.

In Ref.~\cite{Guo:2009wr} a non--relativistic effective
field theory (NREFT) was described, that allows one to
study the role of charmed meson loops in charmonium decays
in a systematic way. Applied to the reaction $\psi'\to
J/\psi\pi^0(\eta)$, the formalism revealed that the contribution
from charmed meson
loops
 is enhanced by a factor of $1/v$, with
$v\simeq0.5$ being the charmed meson velocity\footnote{Here the
velocity is defined via the analytic continuation of the standard
definition, namely $v=\sqrt{-E/M_D}$, with $E$ measured relative
to the open charm threshold.}, compared to the
tree-level one.
In this letter we apply NREFT to
the decays $\psi'\to h_c\pi^0$ and  $\eta_c'\to\chi_{c0}\pi^0$. We
will show that in these two decays, the loop contributions are
highly suppressed, and hence the tree-level terms dominate the decay
amplitudes. Testing experimentally the predictions that emerge here,
especially for the partial decay width for  $\eta_c'\to \chi_{c0} \pi^0$,
would provide a non--trivial test of the NREFT  and is thus
of high importance towards an understanding of the properties of charmonia.

In this context it is instructive to compare the $\psi'\to J/\psi \pi^0$ decays
to  $\psi'\to h_c\pi^0$ on the basis of power counting.  Here we follow the
reasoning of Ref.~\cite{Guo:2009wr}. For the former decay, which happens in a
$p$--wave, the tree level amplitude scales as $m_qq$, where $q$ denotes the
momentum of the final particles in the $\psi'$ rest-frame and a quark mass
factor appears since the reaction is isospin violating. For the same reaction
the loops on the other hand scale as $v^3/v^4(m_q/v^2)[qv^2]=qm_q/v$, where the
the factor $v^3$ comes from the the non--relativistic integral measure and the
$1/v^4$ from the two non--relativistic two--meson propagators. Further, the term
in the round brackets emerges, since pulling out a factor of $m_q$, which is an
energy scale, has to be balanced by a factor which characterizes the intrinsic
energy, $v^2$.  In addition, the $[qv^2]$ term contains the vertex factors from
the external pion coupling, $q$, and from the two $p$--wave vertices in the
loop. Thus, heavy meson loops are enhanced by a factor $1/v\sim 2$.  For the
$s$--wave decay $\psi'\to h_c\pi^0$ on the other hand, the tree level scales as
$m_q$, while the loop here scales as $v^3/v^4(m_q/v^2)[q/M_D]^2$. All factors
that appear are analogous to those discussed above up to the factor $q^2/M_D^2$.
The origin of this is, on the one hand, that the pion is produced in a $p$-wave
and, on the other hand, that there is now only one $p$--wave vertex in the loop
(the $\psi'$ decay) giving rise to a momentum factor at that vertex. To obtain a
non--vanishing result, however, that momentum has to be proportional to $\vec q$
(c.f. Eq.~(\ref{loop}) below). The factor $1/M_D^2$ is then introduced to match
dimensions. Thus, for the reaction $\psi'\to h_c\pi^0$ loops appear to be
suppressed kinematically, on the amplitude level, by a factor
$q^2/(M_D^2v^3)\sim 1/30$. In case of the reaction $\eta_c'\to \chi_{c0} \pi^0$,
we find from the same analysis a suppression of the loops by factor 1/10 ---
here the suppression is weaker due to a larger phase space.  Thus we find on the
basis of the same power counting that heavy meson loops are enhanced compared to
the tree level amplitudes in the isospin violating $p$--wave decays like
$\psi'\to J/\psi \pi^0$, while they are strongly suppressed in $s$--wave decays
like $\psi'\to h_c\pi^0$. The latter observation allows us to predict the
partial decay width of $\eta_c'\to \chi_{c0} \pi^0$. A confirmation of this
prediction would at the same time provide a strong support for the analysis of
Ref.~\cite{Guo:2009wr}.

The $h_c(^1P_1)$ has been the last charmonium state below the $D\bar{D}$
threshold that was confirmed experimentally; for a comprehensive review of the
charmonium physics, see Ref.~\cite{bes-iii}. It was first established in
$p\bar p$ annihilation by the E760 Collaboration at Fermi Lab in
1992~\cite{Armstrong:1992ae}. With $J^{PC}=1^{+-}$, this state cannot be
produced in $e^+ e^-$ annihilation directly. Furthermore, due to the phase
space restriction, it cannot be accessed by  $\psi^\prime$ decays into
$h_c\eta$. Instead, the only open strong decay to $h_c$ is via the
isospin-violating $\psi^\prime\to h_c\pi^0$ process. As a consequence, this branching
ratio is strongly suppressed. Recently, the CLEO-c Collaboration
with $24.5\times 10^6 \ \psi^\prime$ events succeeded in measuring precisely
the product of two branching ratios ${\cal B}(\psi^\prime\to\pi^0 h_c)\times
{\cal B}(h_c\to \gamma\eta_c)=(4.19\pm0.32\pm0.45)\times 10^{-4}$
in~\cite{Dobbs:2008ec}. This combined branching ratio was confirmed by the
BES-III Collaboration~\cite{ligQNP} with 110~million $\psi^\prime$ events in
the same channel, ${\cal B}(\psi^\prime\to\pi^0 h_c)\times {\cal B}(h_c\to
\gamma\eta_c)=(4.58\pm 0.40\pm0.50)\times 10^{-4}$. Furthermore, they reported
the first measurement of the absolute value of the branching ratio for the
$\psi'\to h_c\pi^0$ as ${\cal B}(\psi^\prime\to\pi^0 h_c) = (8.4\pm 1.3\pm
1.0)\times 10^{-4}$ \cite{BESIIIhc}. Using the PDG value for the total width
of the $\psi'$, $\Gamma(\psi') = 309\pm9$~keV~\cite{Amsler:2008zzb},
the partial width of the $\psi'\to h_c\pi^0$ is
\be%
\label{eq:data} \Gamma(\psi'\to h_c\pi^0)=0.26\pm0.05~{\rm keV}.
\ee%

This experimental progress makes it possible to study physics of the $h_c$
to some extent of precision. In particular, its production
in the decay $\psi^\prime\to h_c\pi^0$ appears to be an ideal channel for
investigating the isospin-violating mechanisms and the pertinent
non-perturbative QCD dynamics.  The reaction $\psi^\prime\to
h_c\pi^0$ was first studied theoretically more than thirty years ago
using $\pi^0-\eta$ mixing~\cite{Segre:1976wj}. In order to get an
estimate of the width, the authors estimated the coupling of the
$\psi'$ to the $h_c$ and $\eta$ by assuming it to be equal  to  the
$\psi'J/\psi\eta$ coupling, which was extracted from the measured
$\psi'\to J/\psi\eta$ width. As a result, they obtained $5\ldots
30$~keV for the partial decay width of the $\psi^\prime\to
h_c\pi^0$, which overshoots the measurement significantly.

The QCD  multipole expansion (QCDME) was applied to this
problem~\cite{Eichten:1979ms,Yan:1980uh,Kuang:1981se,Kuang:2002hz},
and the following typical expression was obtained
 \be%
\Gamma[\psi^\prime\to h_c \pi^0] = 0.12\frac {\alpha_M} {\alpha_E} \
\mbox{keV}
\ee%
 where $\alpha_E$ and
$\alpha_M$ are the coupling constants for the color electric dipole
and magnetic dipole gluon radiation, respectively. The
phenomenological determination of the ratio $ {\alpha_M}/{\alpha_E}$
has large uncertainties. By taking the ratio in a range of $1\ldots
3$, the partial width is about $0.12\ldots 0.36$ keV, consistent with
the experimental result. In contrast, a later
calculation~\cite{Ko:1994nw} gives a larger value of 0.84~keV, and
the estimate by Voloshin gives a much smaller value
of~15~eV~\cite{Voloshin:2007dx}. Suffering from a poor knowledge of
the coupling constants~\cite{voloshin-1986,Voloshin:2007dx}, the
QCDME results should rather be regarded as an order-of-magnitude
estimate.

In this work, we shall further investigate the isospin violation
mechanisms of $\psi'\to h_c\pi^0$ and its analogue, the reaction
$\eta_c'\to \chi_{c0}\pi^0$. In Section~\ref{sec:tree}, we will give
the tree level decay amplitudes by constructing the effective chiral
Lagrangian. In Section~\ref{sec:loop} we present the
NREFT~\cite{Guo:2009wr} analysis for the heavy meson loops. As we
will demonstrate the explicit calculation supports the scale
arguments presented above, that heavy meson loops are highly
suppressed for the reactions under consideration. We also checked
that the results are consistent with a calculation using the
ELA~\cite{Li:2008xm,Li:2007au,Zhang:2009kr} --- details will be
presented elsewhere~\cite{Guo:full-length}.  In
Section~\ref{sec:results}, the results for the decay widths are
given. In particular, the width of the $\eta_c'\to \chi_{c0}\pi^0$
is predicted. Some discussions and a summary are given in the last
section.

\section{Tree level contribution}
\label{sec:tree}

Since the mass difference between the initial and final charmonia is
small, the emitted pion is soft. Hence one can construct an effective
chiral Lagrangian considering the charmonia as matter fields. Since
the charmonia are isoscalar and the pion has isospin one, the
transitions violate isospin symmetry. Isospin breaking has two
sources. One is the mass difference between the up and down quarks,
and the other one is of e.m. origin. For the transitions considered
here, the e.m. effect can be neglected (for details, see~\cite{Guo:full-length}).
Defining $\chi=2B_0\cdot{\rm diag}\left(m_u,m_d\right)$, the quark
mass difference is contained in the operator $\chi_-$ which contains an
odd number of pion fields,
\be%
\chi_- = u^\dag \chi u^\dag - u \chi^\dag u,
\ee%
where $B_0= |\langle 0 |\bar q q |0\rangle|/F_\pi^2$, $F_\pi$ is the
pion decay constant in the chiral limit, and $u$ parameterizes the
pion fields as the Goldstone bosons of the spontaneously broken
SU(2)$_L\times$SU(2)$_R$,
\bea%
u = \exp\left(\frac{i\phi}{\sqrt{2}F_\pi}\right), \quad  \phi =
  \left(
    \begin{array}{c c}
 \frac{\pi^0}{\sqrt{2}} & \pi^+\\
\pi^- & - \frac{\pi^0}{\sqrt{2}}
    \end{array}
\right).
\eea%
Using the two-component notation of Ref.~\cite{Hu:2005gf}, the field
for the $S$-wave charmonia $\psi'$ and $\eta_c'$ reads
\be%
J'=\vec{\psi'}\cdot\vec{\sigma}+\eta_c',
\ee%
with $\vec{\psi'}$ and $\eta_c'$ annihilating the $\psi'$ and
$\eta_c'$ states, and $\vec{\sigma}$ the Pauli matrices. The field
for the $P$-wave charmonia is
\be%
\chi^i = \sigma^j
\left(-\chi_{c2}^{ij}-\frac{1}{\sqrt{2}}\epsilon^{ijk}\chi_{c1}^k +
\frac{1}{\sqrt{3}}\delta^{ij}\chi_{c0} \right) + h_c^i,
\ee%
where $\chi_{c2}^{ij}$, $\chi_{c1}^i$, $\chi_{c0}$ and $h_c$
annihilate the $\chi_{c2}$, $\chi_{c1}$, $\chi_{c0}$ and $h_c$
states, respectively.

The leading order (LO) chiral Lagrangian for the transitions
$\psi'\to h_c\pi^0$ and $\eta_c'\to\chi_{c0}\pi^0$ reads
\be%
\label{eq:Ltree} {\cal L} = \frac{i}{4}C \left[
\lang\vec{\chi}^\dag\cdot\vec{\sigma}J'\rang + \lang
J'\vec{\sigma}\cdot\vec{\chi}^\dag\rang \right]
\left(\chi_-\right)_{aa} + {\rm h.c.},
\ee%
where $C$ is an unknown coupling constant, $\lang \cdots \rang$ is the trace in
spinor space, the subscript $a=u,d$ is a flavor index, and repeating indices
$aa$ means evaluating the trace in flavor space. The spin symmetry is violated
due to the presence of the Pauli matrices between the two heavy quarkonium
fields.  Note that the Lagrangian was first proposed in
Ref.~\cite{Casalbuoni:1992fd} in four-component notation, and the coupling
constant $C$ here is $-4/B_0$ times the one defined in that paper. Working out
the traces in both the spinor and flavor space, one finds that only the two
transitions considered in this paper are allowed as dictated by
 conservation of angular momentum,  parity and charge conjugation
invariance,
\be%
\label{eq:Ltreeexpansion} {\cal L} = i C \left(\vec{\psi}'\cdot
\vec{h_c}^\dag + \sqrt{3} \eta_c'\chi_{c0}^\dag \right)
\left(\chi_-\right)_{aa} + {\rm h.c.},
\ee%
and, after taking into account the $\pi^0-\eta$ mixing which contributes to the
isospin breaking transitions,
\be%
\left(\chi_-\right)_{aa} = 6i \frac{B_0}{F_\pi} (m_d-m_u) \tilde\pi^0 + \ldots~,
\ee%
where $\tilde\pi^0=\pi^0+\epsilon_{\pi^0\eta}\eta$ is the physical pion field
with $\epsilon_{\pi^0\eta}$ being the $\pi^0-\eta$ mixing angle. In the above
equation, we have neglected the multi-pion terms. Now it is easy to write out
the tree level amplitudes,
\bea%
{\cal M}(\psi'\to h_c\pi^0) \al=\al \sqrt{M_{\psi'}M_{h_c}}\frac{6}{F_\pi}C
%\sqrt{M_{\psi'}M_{h_c}}
\vec{\varepsilon}\,(\psi')\cdot\vec{\varepsilon}\,(h_c) \non\\
\al\al \times B_0(m_d-m_u),
\non\\
{\cal M}(\eta_c'\to \chi_{c0}\pi^0) \al=\al
\sqrt{M_{\eta_c'}M_{\chi_{c0}}}\frac{6\sqrt{3}}{F_\pi}C
%\sqrt{M_{\eta_c'}M_{\chi_{c0}}}
B_0(m_d-m_u), \label{eq:Amptree}
\eea%
Note that the above amplitudes were multiplied by a factor
$\sqrt{M_fM_i}$, with $M_{i(f)}$ being the mass of the initial
(final) charmonium, to account for the non-relativistic
normalization of the heavy fields in the Lagrangian.

\section{Charmed meson loops}
\label{sec:loop}

\begin{figure}
\begin{center}
\vglue3mm \includegraphics[width=0.5\textwidth]{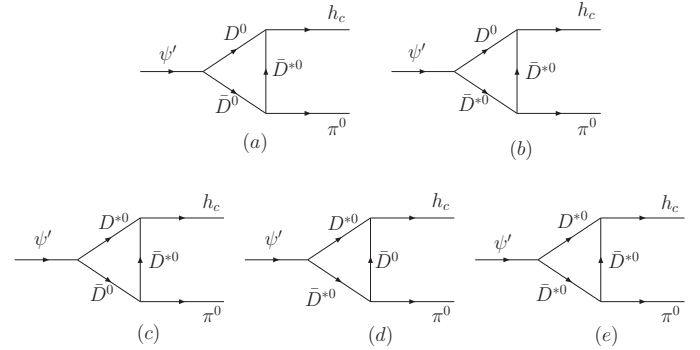}
 \caption{Diagrams of neutral intermediate meson loops contributing to $\psi^\prime\to
 h_c\pi^0$. Their charge conjugate diagrams are implied. } \protect\label{fig-1}
\end{center}
\end{figure}

In this section we investigate the contribution of charmed meson
loops to the two decays. The relevant diagrams for the
reaction  $\psi'\to h_c\pi^0$ with neutral intermediate states are shown in
Fig.~\ref{fig-1}.
The coupling of pion to the charmed mesons is described by heavy
meson chiral perturbation
theory~\cite{Burdman:1992gh,Wise:1992hn,Yan:1992gz} (for a review,
see Ref.~\cite{Casalbuoni:1996pg}). The fields for the pseudoscalar
and vector charmed mesons in the same spin multiplet can be written
as $H_a=\vec{V}_a\cdot\vec{\sigma}+P_a$ with $V_a$ and $P_a$
denoting the vector and pseudoscalar charmed
mesons~\cite{Hu:2005gf}, respectively, where $a$ is the flavor index
with $\{ P_u, P_d\} = \{ D^0, D^+ \}$ and similar for the vector
mesons. The LO chiral effective Lagrangian for the axial coupling
is~\cite{Hu:2005gf}
\be%
{\cal L_\phi} = -{g\over 2} \lang
H_a^{\dag}H_b\vec{\sigma}\cdot\vec{u}_{ba}\rang, \label{eq:Lphi}
\ee%
where the axial current is
$\vec{u}=-\sqrt{2}\vec{\partial}\phi/F_\pi +{\cal O}(\phi^3)$, and
$g$ the pertinent coupling constant.

The LO Lagrangian for the coupling of the $S$- or $P$-wave charmonium
fields to the charmed and anti-charmed mesons can be constructed
respecting parity, charge conjugation and spin symmetry. The one for the
$\psi'$ and $\eta_c'$ reads~\cite{Guo:2009wr}
\be%
{\cal L}_\psi = i \frac{g_2'}{2} \lang J'^\dag H_a
\vec{\sigma}\cdot\!\stackrel\leftrightarrow{\partial}\!{\bar
H}_a\rang + {\rm h.c.}, \label{eq:Lpsi0}
\ee%
where $A\!\stackrel\leftrightarrow{\partial}\!\!B\equiv
A(\vec{\partial}B)-(\vec{\partial}A)B$, and ${\bar H}_a=-\vec{{\bar
V}}_a\cdot\vec{\sigma}+{\bar P}_a$ is the field for anti-charmed
mesons~\cite{Fleming:2008yn}. The LO Lagrangian for the $P$-wave
charmonia spin multiplet is~\cite{Fleming:2008yn}
\be%
{\cal L}_\chi = i \frac{g_1}{2} \lang\chi^{\dag i} H_a \sigma^i
{\bar H}_a\rang + {\rm h.c.}. \label{eq:Lchi0}
\ee%
These Lagrangians were introduced in Ref.~\cite{Colangelo:2003sa} in
four-component notation. The values of the coupling constants $g_1$ and $g_2$ in
that paper are half of those introduced here.

Using these Lagrangians, one can work out the decay amplitudes. For
simplicity, we focus on diagram (a) in Fig.~\ref{fig-1} --- the
analysis of the other diagrams is analogous. The amplitude for
$\psi'\to h_c\pi^0$ from the diagram (a) reads (for the full
amplitudes and calculation details, we refer to
Ref.~\cite{Guo:full-length})
\begin{widetext}
\bea%
{\cal M}(\psi'\to h_c\pi^0)_{\rm (a)} = 2\frac{g}{F_\pi}g_1g_2'
\vec{q}\cdot\vec{\varepsilon}(h_c) \vec{q}\cdot\vec{\varepsilon}(\psi') R
%\non\\\al\al \times%
\left[I^{(1)}(q,D^0,D^0,D^{*0}) - I^{(1)}(q,D^\pm,D^\pm,D^{*\pm})\right],
\eea%
\end{widetext}
where $q$ is the pion three-momentum in the rest-frame of the
initial charmonium, $R=\sqrt{M_{\psi'}M_{h_c}}M_D^2M_{D^*}$ is
included to account for the non-relativistic normalization of the
heavy fields, and we have labeled the loop function $I^{(1)}(q)$ by
the intermediate charmed mesons. It is obvious that the isospin
breaking comes from the neutral and charged meson mass differences.
The loop function, evaluated in the rest-frame of
the decaying particle, is defined by
\begin{widetext}
\bea%
q^i I^{(1)}(q) \al=\al  \frac{i}{8m_1m_2m_3}
 \int\!\frac{d^4l}{(2\pi)^4} \frac{l^i}{
\left(l^0{-}T_1(|\vec l|)\right) \left(P^0{-}l^0{-}T_2(|\vec
l|)\right)
\left(l^0{-}q^0{-}T_3(|\vec l{-}\vec q|)\right) } \non\\
 \al=\al  \frac{1}{8m_1m_2m_3}
 \int\!\frac{d^3l}{(2\pi)^3} \frac{l^i}{
\left(E_i{-}T_2(|\vec l|){-}T_1(|\vec l|)\right)
\left(E_f{-}T_2(|\vec l|){-}T_3(|\vec l{-}\vec q|)\right) }
\label{loop}
\eea%
\end{widetext}
where $T_i(p)=p^2/2m_i$ denotes the kinetic energy for the charmed
mesons with masses $m_1,m_2$ and $m_3$, $E_i=M_i-m_1-m_2$ and
$E_f=M_f-m_2-m_3-E_\pi$ denote the energies available for the first
(before pion emission) and second (after pion emission) two heavy
meson intermediate state.
 For diagram (a),
$m_1=m_2=M_{D}$, $m_3=M_{D^*}$, $M_i=M_{\psi'}$, and $M_f=M_{h_c}$. The loop
function is convergent~\cite{Guo:2009wr,Guo:full-length}. Defining
$c=2\mu_{12}b_{12}$ and $c'=2\mu_{23}b_{23}$, with $\mu_{ij}=m_im_j/(m_i+m_j)$
being the reduced mass, $b_{12}=m_1+m_2-M_{\psi'}$ and
$b_{23}=m_2+m_3+E_{\pi}-M_{\psi'}$, the loop function can be approximated by
\be%
I^{(1)}(q) = N \frac{\mu_{23}}{m_3}
\frac{2\left(\sqrt{c'}+2\sqrt{c}\right)}{3\left(\sqrt{c'}+\sqrt{c}\right)^2}
\ee%
with $N=\mu_{12}\mu_{23}/(16\pi m_1m_2m_3)$, where terms of order $\vec{q}\
^2/c'$ and higher have been neglected. The approximation is reasonable because
for either of the two decays considered here, the pion momentum is small and
fulfills $\vec{q}\ ^2\ll c'$.

We can now have another, more refined look, at the order of magnitude estimate
for the loop function. Since $\sqrt{c}$ and $\sqrt{c'}$ are approximately the
momenta of the charmed mesons in the loop, we count them as $M_Dv$ with $v$
being the velocity of the charmed mesons. It follows that $b_{12}\sim b_{23}\sim
M_D v^2$. For an order-of-magnitude estimate, one may neglect the difference
between $c$ and $c'$, and denoting them by $2\mu b$. Then, $I^{(1)}(q)\sim
N/(4\sqrt{2\mu b})$. Denoting the mass difference between the charged and
neutral charmed mesons by $\delta$, we have $\mu_c=\mu_n+\delta/2$ and
$b_c=b_n+2\delta$, where the lower-index $n(c)$ means neutral (charged). The
amplitude for diagram (a) scales as
\bea%
{\cal M}(\psi'\to h_c\pi^0)_{\rm (a)} \al\sim\al N_{\rm (a)}
\vec{q}\ ^2 \frac{N}{4} \left( \frac{1}{\sqrt{2\mu_nb_n}} -
\frac{1}{\sqrt{2\mu_cb_c}} \right) \non\\
\al\sim\al N_{\rm (a)} \frac{N}{4} \delta v \frac{\vec{q}\ ^2}{b_n^2} \non\\
\al\sim\al N_{\rm (a)} \frac{N}{4} \delta \frac{1}{v^3}
\frac{\vec{q}\ ^2}{M_D^2},
\eea%
where $N_{\rm (a)}=2(g/F_\pi)g_1g_2'R\vec{q}\cdot\vec{\varepsilon}(h_c)
\vec{q}\cdot\vec{\varepsilon}(\psi') / \vec{q}\ ^2$. Thus, we confirm the
parametric behavior derived in the introduction on the basis of the NREFT. For
the transition from the $\psi'$ to the $h_c$, the pion momentum in the $\psi'$
rest-frame is $q=q_1=86$~MeV, and hence $\vec{q}\ ^2/M_D^2\simeq 2\times
10^{-3}$. Taking into account that the velocity $v$ may be roughly estimated as
$\sqrt{[2M_{\hat{D}}-(M_{\psi'}+M_{h_c})/2]/M_{\hat{D}}}\simeq0.4$ with
$M_{\hat{D}}$ being the averaged charmed meson mass, the dimensionless factor
\be%
\frac{1}{v^3} \frac{\vec{q}\ ^2}{M_D^2} \simeq 0.03
\ee%
produces a significant suppression compared to the tree level amplitude.
 A similar
though more moderate suppression happens in case of the $\eta_c'\to
\chi_{c0}\pi^0$ also. For this decay, the momentum of the pion is
$q=q_2=171$~MeV, and the suppression factor is
\be%
\frac{1}{v^3} \frac{\vec{q}\ ^2}{M_D^2} \simeq 0.1.
\ee%

The  numerical results support the above power counting
argument. If we only consider the contribution from the
charmed-meson loops, the widths of the $\psi'\to h_c\pi^0$ and
$\eta_c'\to \chi_{c0}\pi^0$ are
\bea%
\label{eq:psi'hc_loop}
\Gamma(\psi'\to h_c\pi^0)_{\rm loop} \al=\al
2.1\times10^{-7} g_1^2 g_2^{\prime\ 2}~{\rm keV}, \non\\
\Gamma(\eta_c'\to \chi_{c0}\pi^0)_{\rm loop} \al=\al 1.0\times10^{-5} g_1^2
g_2^{\prime\ 2}~{\rm keV},
\eea%
where the $\pi^0-\eta$ mixing has been taken into account, and the values of
$g_1$ and $g_2'$ are given in units of GeV$^{-1/2}$ and GeV$^{-3/2}$,
respectively. We checked that the ELA gives similar results, which confirms our
analysis.

One may ask if the values of the coupling constants are so large that the
suppression gets invalidated. In fact, because all the charmonia considered here
are below the $D{\bar D}$ threshold, the couplings cannot be extracted directly
from the decay widths. However, one may get a feeling about their values from
other sources or from model calculations. Assuming the coupling of the $\psi'$
to the charmed mesons has similar strength as the one of the $J/\psi$, we obtain
$g_2'\simeq2$~GeV$^{-3/2}$ from the $\psi'\to J/\psi\pi^0(\eta)$ where the
charmed meson loops dominate~\cite{Guo:2009wr}. Values of the same order of
magnitude were obtained from various model calculations, see e.g.
Refs.~\cite{Colangelo:2003sa,Haglin:2000ar,Deandrea:2003pv}. In
Ref.~\cite{Colangelo:2003sa}, the authors estimated $g_1$ using vector meson
dominance, which gives $g_1=-4.2$~GeV$^{-1/2}$. Using these values, the
resulting width $\Gamma(\psi'\to h_c\pi^0)_{\rm loop} \simeq 1\times 10^{-5}$~
keV. It is smaller by 4 orders of magnitude compared to  the BES-III
measurement, and it confirms our power counting estimate presented before.

In addition, even without any assumption on the coupling constants,
from our analysis we can predict
\begin{equation}
\frac{\Gamma(\eta_c'\to \chi_{c0}\pi^0)_{\rm loop}}{\Gamma(\psi'\to h_c\pi^0)_{\rm loop}} =
48 \ .
\end{equation}
As we will see, this ratio as derived solely from the loop contributions,
is much larger than the corresponding one derived from including the
tree level amplitudes only. Note, however, that because $v\simeq 0.5$,
there might be sizeable corrections to this result.
Thus, for the mentioned decays it can be
tested experimentally, if there is a dominance from the loops or from
the tree level contribution.

\section{Decay widths}
\label{sec:results}

As shown in the last section, the charmed meson loops can be
neglected. Hence, the LO decay amplitudes are given by
Eq.~(\ref{eq:Amptree}). Then the decay widths are
\bea%
\Gamma\left(\psi'\to h_c\pi^0\right) \al=\al \frac{q_1 M_{h_c}}{8\pi
M_{\psi'}} C^2
\left[\frac{6}{F_\pi}B_0(m_d-m_u)\right]^2, \non\\
\Gamma\left(\eta_c'\to\chi_{c0}\pi^0\right) \al=\al \frac{3q_2
M_{\chi_{c0}}}{8\pi M_{\eta_c'}} C^2 \left[\frac{6}{F_\pi}B_0(m_d-m_u)\right]^2.
\eea%
The pion momenta $q_i$ where introduced in the previous section.
 The ratio of these two
widths is free of any parameter
\be%
\frac{\Gamma\left(\eta_c'\to\chi_{c0}\pi^0\right)}{\Gamma\left(\psi'\to
h_c\pi^0\right)} = 3
\frac{q_2}{q_1}\frac{M_{\chi_{c0}}M_{\psi'}}{M_{\eta_c'}M_{h_c}} =
5.86\pm0.94,
\ee%
where the 15\% uncertainty comes from neglecting higher order terms
in either the heavy quark expansion or the chiral expansion
$${\cal O}\left(\frac{\Lambda_{\rm QCD}}{m_c}\right) \sim {\cal O}\left(\frac{m_\pi}{\Lambda_\chi}\right) \sim 15\%,$$
where $\Lambda_\chi\simeq1$~GeV, as well as heavy meson loops.
 Using the experimental value of
$\Gamma\left(\psi'\to h_c\pi^0\right)$, we predict the width of
the $\eta_c'\to\chi_{c0}\pi^0$ as
\be%
\Gamma\left(\eta_c'\to\chi_{c0}\pi^0\right) = 1.5\pm0.3\pm0.2~{\rm
keV},
\ee%
where the first uncertainty is experimental and the second
theoretical due to neglecting higher orders. With the total width of
the $\eta_c'$, $\Gamma (\eta_c' ) =
14\pm7$~MeV~\cite{Amsler:2008zzb}, the branching
fraction of the isospin-breaking transition is
\be%
{\cal B}\left(\eta_c'\to\chi_{c0}\pi^0\right) = (1.1\pm0.6)\times
10^{-4}.
\ee%
Note, as a contrast, if the reactions were dominated by heavy meson
loops, the predicted branching fraction would be larger by a factor
of about 5. The prediction, and therefore the dynamics underlying
the decays, is testable with $\overline{\rm P}$ANDA at
FAIR~\cite{Lutz:2009ff}.

Furthermore, one may check if the tree-level contribution gives the
right order of magnitude of the decay width $\Gamma\left(\psi'\to
h_c\pi^0\right)$ through dimensional analysis.
The tree--level amplitudes of Eq.~(\ref{eq:Amptree})
are proportional to the dimensionless factor $\sqrt{M_fM_i}C$.
Because the spin symmetry is violated as can be
seen from the presence of the Pauli matrices in
Eq.~(\ref{eq:Ltree}), one may write
\be%
\label{eq:Cestimate} \sqrt{M_fM_i}C = \tilde{C} \frac{\Lambda_{\rm QCD}}{m_c}
\ ,
\ee%
with the dimensionless parameter $\tilde{C}$ being a number of
natural size, i.e. of order one.
Using the current knowledge of the quark mass
ratio~\cite{Leutwyler:Stern}
\be%
r\equiv \frac{m_u}{m_d}=0.47\pm0.08,
\ee%
and the LO relation between the pion mass and the quark masses
$m_{\pi^0}^2=B_0(m_u+m_d)$ (neglecting strong isospin violation),
we get
\be%
\label{eq:Bmdu} B_0(m_d-m_u) = \left(\frac{1-r}{1+r}\right) m_{\pi^0}^2 =
(6.6\pm2.0)\times10^{-3}~{\rm GeV}^2,
\ee%
where the uncertainty is dominated by that of the quark mass ratio.
Using the value given in
Eq.~(\ref{eq:Bmdu}), the width for the $\psi'\to h_c\pi^0$ is
\be%
\Gamma\left(\psi'\to h_c\pi^0\right) = (0.9\pm0.6)\tilde{C}^2~{\rm keV}.
\ee%
Since $\tilde{C}$ is of order 1, the above result agrees with the BES-III
measurement $0.26\pm0.05$~keV well. The agreement in turn supports the
tree-level dominance argued for in this paper. It is also consistent with the
QCDME results reported in Refs.~\cite{Kuang:2002hz,Ko:1994nw}.

\section{Discussion and summary}
\label{sec:sum}

We demonstrated that, based on a non--relativistic effective field theory (NREFT),
charmed meson loops are highly suppressed by a factor
$v\vec{q}\ ^2/(M_Dv^2)^2\ll 1$ in  $\psi'\to h_c\pi^0$ and $\eta_c'\to
\chi_{c0}\pi^0$, which are transitions between one $P$-wave and one $S$-wave
charmonia. The reason for the suppression is that, due
to the small phase spaces of the two transitions, the pion momentum is much
smaller than the approximate kinetic energy of the intermediate charmed
mesons. The situation is completely different for the transitions
between two $S$-wave charmonia. For the $\psi'\to J/\psi\pi^0(\eta)$, the
charmed meson loops are enhanced by a factor of $1/v$ compared with the tree
level contribution~\cite{Guo:2009wr}. There is no factor proportional to
$\vec{q}\ ^2$, and hence the relative size of the pion momentum to the kinetic
energy of the virtual charmed mesons does not have an impact. The difference in
these two cases is a consequence of the difference in the quantum numbers of
the $J/\psi$ and $h_c$ which determine their coupling to the charmed mesons.

We note in passing that it follows from our findings that
approximating the $\psi'h_c\eta$ coupling by that for
$\psi'J/\psi\eta$, as done in Ref.~\cite{Segre:1976wj}, is not
justified. In the light of this one then understands why the estimate made there is much
larger than the measurement.

In summary, we have shown that intermediate charmed meson loops are highly
suppressed in the decays $\psi'\to h_c\pi^0$ and $\eta_c'\to \chi_{c0}\pi^0$,
which is completely different from the situation of the $\psi'\to
J/\psi\pi^0(\eta)$. We confirmed the general power counting arguments given by
explicit calculations both within the NREFT as well as an effective Lagrangian
approach. By constructing the LO chiral Lagrangian for the decays, and employing
the experimental result for $\Gamma(\psi'\to h_c\pi^0)$, we give a
model-independent prediction for the width of the $\eta_c'\to \chi_{c0}\pi^0$,
which can be measured at $\overline{\rm P}$ANDA, $\Gamma (\eta_c'\to \chi_{c0}
\pi^0) = 1.5\pm0.4$~keV, where the experimental and theoretical uncertainties
have been summed in quadrature. Note, were the transitions dominated by the
heavy meson loop contributions, the predicted partial decay width would be
larger by  a factor of about 8. An experimental confirmation of our prediction
would provide a strong support for the NREFT employed. Having available an
effective field theory that allows one to study both direct transitions as well
as those mediated via heavy loops is an important step towards a detailed
theoretical understanding of charmonium states.

\vspace{5mm}
\section*{Acknowledgments}
Q.Z. and G.L. acknowledge the supports, in part, from the National Natural
Science Foundation of China (Grants No. 10675131 and 10491306), Chinese Academy
of Sciences (KJCX3-SYW-N2), and Ministry of Science and Technology of China
(2009CB825200). The work of F.K.G., C.H. and U.G.M. is supported by  the
Helm\-holtz Association through funds provided to the virtual institute ``Spin
and strong QCD'' (VH-VI-231) and by the DFG (SFB/TR 16, ``Subnuclear Structure
of Matter''), and the European Community-Research Infrastructure Integrating
Activity ``Study of Strongly Interacting Matter'' (acronym HadronPhysics2, Grant
Agreement n. 227431) under the Seventh Framework Programme of EU.

\end{document}